\documentclass[cits]{PoS}

\usepackage{bm}

\title{Quarkonium mass splittings with Fermilab heavy quarks and 2+1
  flavors of improved staggered sea quarks }

\ShortTitle{Quarkonium with Fermilab quarks}

\author{T.~Burch\thanks{Present address: 
        Institut f\"ur Theoretische Physik, Universit\"at Regensburg, 93040 Regensburg, Germany.
        },
        C.E.~DeTar and \speaker{L.~Levkova}\\
        Physics Department, University of Utah, Salt Lake City, UT 84112, USA\\
        E-mail: \email{ludmila@physics.utah.edu}}

\author{M.~Di Pierro \\
School of Computer Sci., Telecom.\ and Info.\ Systems, 
DePaul University, Chicago, Illinois, USA}

\author{A.X.~El-Khadra \\
Physics Department, University of Illinois, Urbana, 
Illinois, USA}

\author{Steven~Gottlieb \\
Department of Physics, Indiana University, Bloomington, 
IN 47405, USA}

\author{A.S.~Kronfeld,  P.B.~Mackenzie, and J.N.~Simone \\
Fermi National Accelerator Laboratory, Batavia, IL 60510, 
USA}

\abstract{We present results from an ongoing lattice study of the
  lowest lying charmonium and bottomonium level splittings using the
  Fermilab heavy quark formalism.  Our objective is to test the
  performance of this action on MILC-collaboration ensembles of
  $(2+1)$ flavors of light improved staggered (asqtad) quarks.
  Measurements are done on 16 ensembles with degenerate up and down
  quarks of various masses, thus permitting a chiral extrapolation,
  and over lattice spacings ranging from $0.09$ fm to $0.18$ fm, thus
  permitting study of lattice-spacing dependence.  We examine
  combinations of the mass splittings that are sensitive to components
  of the effective quarkonium potential.}

\FullConference{The XXVII International Symposium on Lattice Field Theory - LAT2009\\
		 July 26-31 2009\\
		 Peking University, Beijing, China}

\newcommand{\bi}{\begin{itemize}}
\newcommand{\ei}{\end{itemize}}
\newcommand{\be}{\begin{equation}}
\newcommand{\ee}{\end{equation}}
\newcommand{\bea}{\begin{eqnarray}}
\newcommand{\eea}{\end{eqnarray}}
\newcommand{\et}{{\it et al.}}
\newcommand{\ie}{{\it i.e.}}

\newcommand{\case}[2]{\ensuremath{{\textstyle\frac{#1}{#2}}}}
\newcommand{\figwidth}{0.9\textwidth}

\begin{document}

\section{Introduction}

The well-studied charmonium and bottomonium systems have long been
used as a test bed for phenomenological models and lattice methods.
It is well known that including light sea quarks is essential for
obtaining good agreement with experiment.  Few studies have carried
out a systematic treatment that includes both the chiral (sea quark)
and continuum limit.  The present study describes progress to date in
such an ongoing study \cite{DiPierro:2002ta}.
It is based on charm and bottom masses that were determined in
previous studies \cite{Bernard:2009}. We limit our attention to
lattices with spacing $a \ge 0.09$ fm. Table \ref{tab:ensembles} lists
the 16 ensembles used in this study
\cite{Bernard:2001av,Aubin:2004wf,Bazavov:2009}.

\begin{table}
\begin{center}
\begin{tabular}{|l|l|l|}
\hline
ensemble & $a$ (approx) (fm) & sea quark ratio $m_{ud}/m_s$ \\
\hline
Extra coarse  & 0.18 &  0.6, 0.4, 0.2, 0.1 \\
Medium coarse & 0.15 &  0.6, 0.4, 0.2, 0.1 \\
Coarse        & 0.12 &  0.6, 0.4, 0.2, 0.15, 0.1 \\
Fine          & 0.09 &  0.4, 0.2, 0.1 \\
\hline
\end{tabular}
\end{center}
\label{tab:ensembles}
\caption{Light quark mass ratios and lattice spacings for the
  ensembles used in this study. The strange quark mass is set
  to approximately its physical value.}
\end{table}

We simulate the heavy charm and bottom quarks with the Fermilab action
\cite{ElKhadra:1996mp},

\begin{eqnarray*}
	S = \sum_n \bar{\psi}_n\psi_n & - & \kappa\sum_n \left[
		\bar{\psi}_n(1-\gamma_4)U_{n,4}\psi_{n+\hat{4}} +
		\bar{\psi}_{n+\hat{4}}(1+\gamma_4)U^\dagger_{n,4}\psi_n 
	\right] \\
	& - & \kappa\zeta \sum_{n,i} \left[
		\bar{\psi}_n(r_s-\gamma_i)U_{n,i}\psi_{n+\hat{\imath}} +
		\bar{\psi}_{n+\hat{\imath}}(r_s+\gamma_i)U_{n,i}^\dagger\psi_n
	\right] \label{eq:S} \\
	& - & c_B\kappa\zeta \sum_n
		\bar{\psi}_ni\bm{\Sigma}\cdot\bm{B}_n\psi_n -
		c_E\kappa\zeta \sum_{n;i} 
		\bar{\psi}_n\bm{\alpha}\cdot\bm{E}_n\psi_n \, \, .
\end{eqnarray*}
The energy of a single quark of spatial momentum ${\bf p}$ in
nonrelativistic approximation is
\begin{displaymath}
	E(\bm{p}) = m_1 + \frac{\bm{p}^2}{2m_2} + O(p^4),
	\label{eq:energy}
\end{displaymath}
where $m_1$ is the rest mass and $m_2$ is the ``kinetic'' mass.  They
can be made equal if we tune the temporal anisotropy $\zeta$.
Instead, we set $\zeta = 1$ and limit our attention to mass splittings
for which the additive mass renormalization cancels.  We also take
$c_B = c_E = 1/u_0^3$, where $u_0$ is the tadpole factor.  These
choices are explained in greater detail in \cite{Burch:2009}.  The
resulting action is just the standard clover action with the clover
coefficient set according to the Fermilab interpretation.

\section{Tuning the heavy quark masses}

There are a variety of possible ways to determine the masses
($\kappa$'s) of the charm and bottom quarks.  Since we know the
lattice scale from other measurements, determining the heavy quark
mass involves matching a lattice mass with an experimentally observed
mass. Tuning to the rest mass $M_1$ of quarkonium is clearly
inaccurate, since it inherits the large additive renormalization of
the quark mass $m_1$.  Tuning the kinetic mass $M_2$ of quarkonium is
a possibility, but that mass includes a strong binding energy that we
would like to study independently of the tuning
\cite{Kronfeld:1996uy}.  So a cleaner approach tunes to the
spin-averaged kinetic masses of the $\bar D_s = \frac{1}{4} m_{D_s} +
\frac{3}{4} m_{D_s^*}$ and the corresponding $\bar B_s$ multiplet
\cite{Bernard:2009,Burch:2009}.  The heavy-light system has only a
mild binding contribution.  In this way our study of quarkonium
binding is more predictive.  Results of tuning are shown in
Table~\ref{tab:tunedkappas}.  Tuning errors are discussed in detail in
Ref.~\cite{Burch:2009}.

\begin{table}[t]
\begin{center}
\begin{tabular}{|l|l|l|l|}
\hline
ensemble       & $a$     & $\kappa_c$ & $\kappa_b$ \\
\hline
Extra coarse   & 0.18 fm & 0.120      & $-$   \\
Medium coarse  & 0.15 fm & 0.122      & 0.076 \\
Coarse         & 0.12 fm & 0.122      & 0.086 \\
Fine           & 0.09 fm & 0.127      & 0.0923 \\
\hline
\end{tabular}
\end{center}
  \label{tab:tunedkappas}
  \caption{Tuned charm and bottom $\kappa$'s.}
\end{table}

All tuning methods should agree in the continuum limit.  Discrepancies
at nonzero lattice spacing come from discretization artifacts that
grow with $ma$, \ie, the quark mass in lattice units.  So, for
example, at $a = 0.15$ fm we find that the tuned charm mass is
approximately the same whether obtained from the kinetic mass of the
$D_s$ multiplet or the kinetic charmonium mass, but the tuned bottom
mass differs significantly: $\kappa_b = 0.94$ from tuning the kinetic
bottomonium mass and $0.76$ from tuning the $B_s$ multiplet.

Nonetheless, there are situations that require tuning the quarkonium
rest mass.  For our companion study of charmonium annihilation
effects, mixing between quarkonium states and glueball states could be
important \cite{Levkova-DeTar:2008}. In this case it is important to
arrange for a correct placement of the unmixed charmonium and glueball
eigenenergies of the lattice hamiltonian, \ie, the unmixed rest
masses \cite{Chen:2006}.  However, in that study we hope for at best
15\% accuracy in computing the tiny mass shifts coming from
annihilation, so we tolerate a mistuning of the kinetic quark mass.

\section{Results}

We measure quarkonium correlators with smeared relativistic and
nonrelativistic S-wave and P-wave sources and sinks.  To extract
masses, we use a multistate fit model with loose Bayesian priors, and
we determine statistical errors in mass splittings from a bootstrap
analysis.  We present a sampling of results.  More are given in
\cite{Burch:2009}.  We examine them in terms of a traditional
nonrelativistic decomposition of the effective heavy quark potential,
namely, central, spin-spin, spin-orbit, and tensor contributions.

\begin{figure}[t]
\begin{center}
\begin{tabular}{cc}
  \begin{minipage}[t]{0.5\textwidth}
\includegraphics[width=\figwidth]{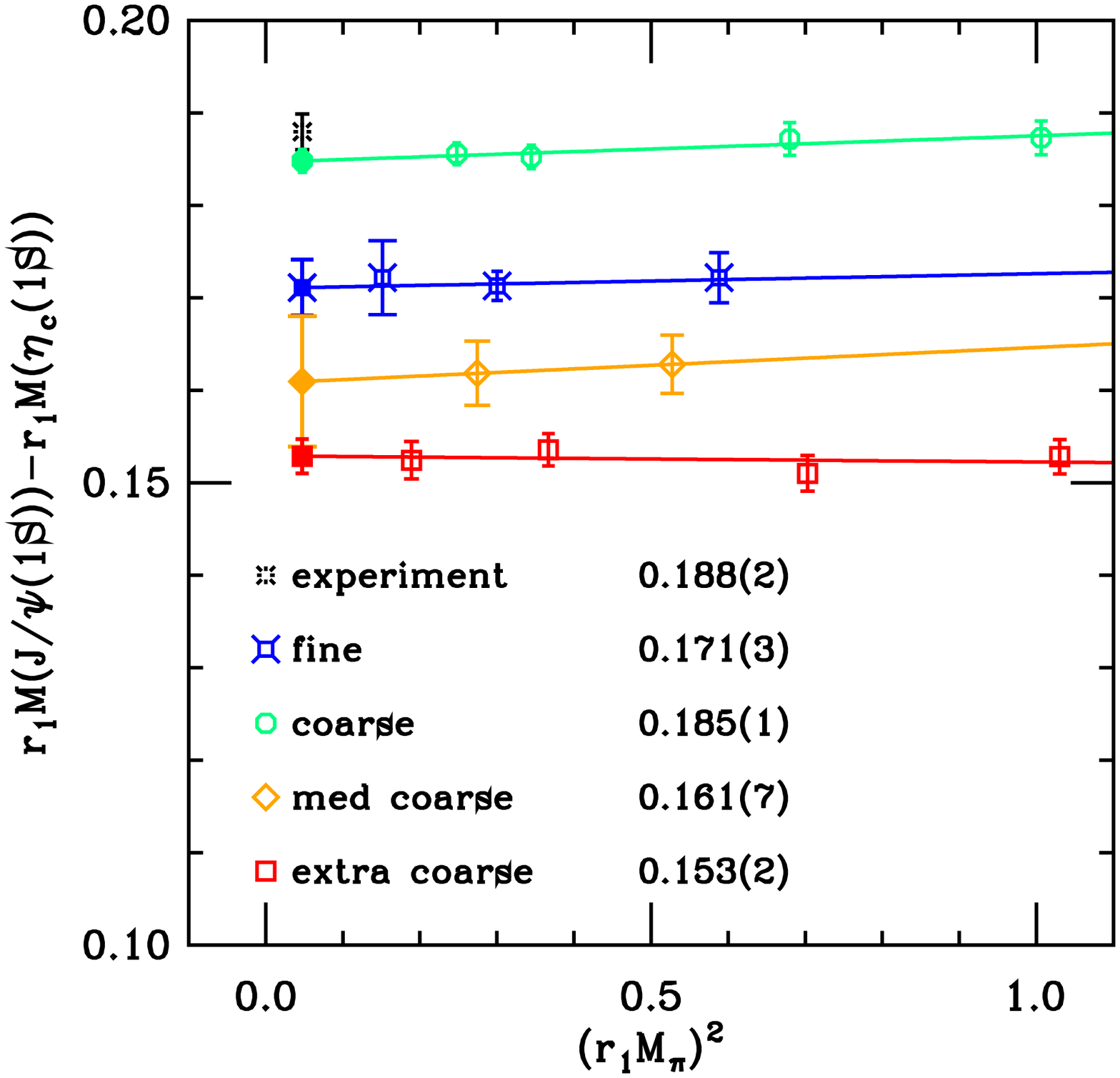} \\
  \end{minipage}
&
  \begin{minipage}[t]{0.5\textwidth}
\includegraphics[width=\figwidth]{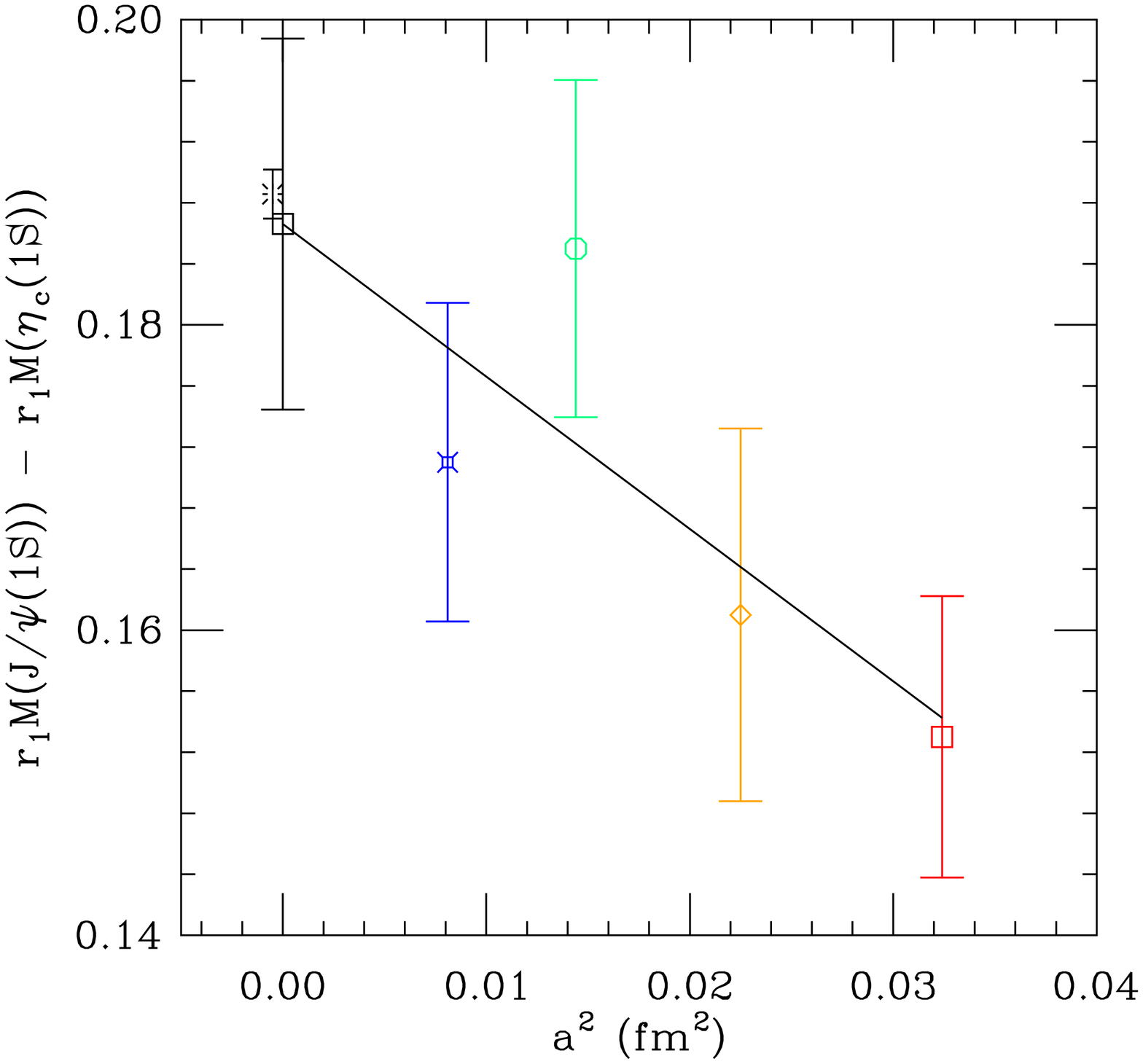}\\
  \end{minipage}
\end{tabular}
\end{center}
\caption{Results for charmonium hyperfine splitting. Splittings are in
  $r_1 = 0.318$ fm units. (1/$r_1 = 620$ MeV). The left panel shows
  the chiral extrapolation with only statistical errors shown.  The
  right panel shows the continuum extrapolation in $a^2$ with kappa
  tuning errors of 6\% included. }
\label{fig:chfs}
\end{figure}

\subsection{Charmonium hyperfine splitting}  

Hyperfine splitting provides a direct measure of the strength of the
spin-spin chromomagnetic interaction.  In Fig.~\ref{fig:chfs} we show
our results for charmonium hyperfine splitting.  Here only the quark
line ``connected'' diagrams are included.  The dependence on sea quark
mass is evidently quite weak. The continuum extrapolation, shown with
kappa tuning errors included, gives 116(5) MeV compared with 117(1)
MeV from experiment.  It would clearly be good to reduce $\kappa$-tuning
errors.

The contribution to the charmonium correlator and mass from quark line
disconnected diagrams is expected to be small, so they are usually
ignored.  Because it is so small, it is a challenge to calculate it
\cite{Levkova-DeTar:2008}.  Our most recent results are given in
Table~\ref{tab:disc}.  We find that annihilation processes actually
decrease the magnitude of the splitting.  The effect is smaller than
or comparable to our current errors in the connected contribution.

\begin{table}
\begin{center}
\begin{tabular}{lll}
\hline
superfine & 0.06 fm & $-3.4(3)$ MeV \\
fine      & 0.09 fm & $-5.5(8)$ MeV \\
\hline
\end{tabular}
\end{center}
\label{tab:disc}
  \caption{Contribution from charm annihilation to the charmonium
    hyperfine splitting}
\end{table}

\begin{figure}[t]
\begin{center}
\begin{tabular}{cc}
  \begin{minipage}[t]{0.5\textwidth}
\includegraphics[width=\figwidth]{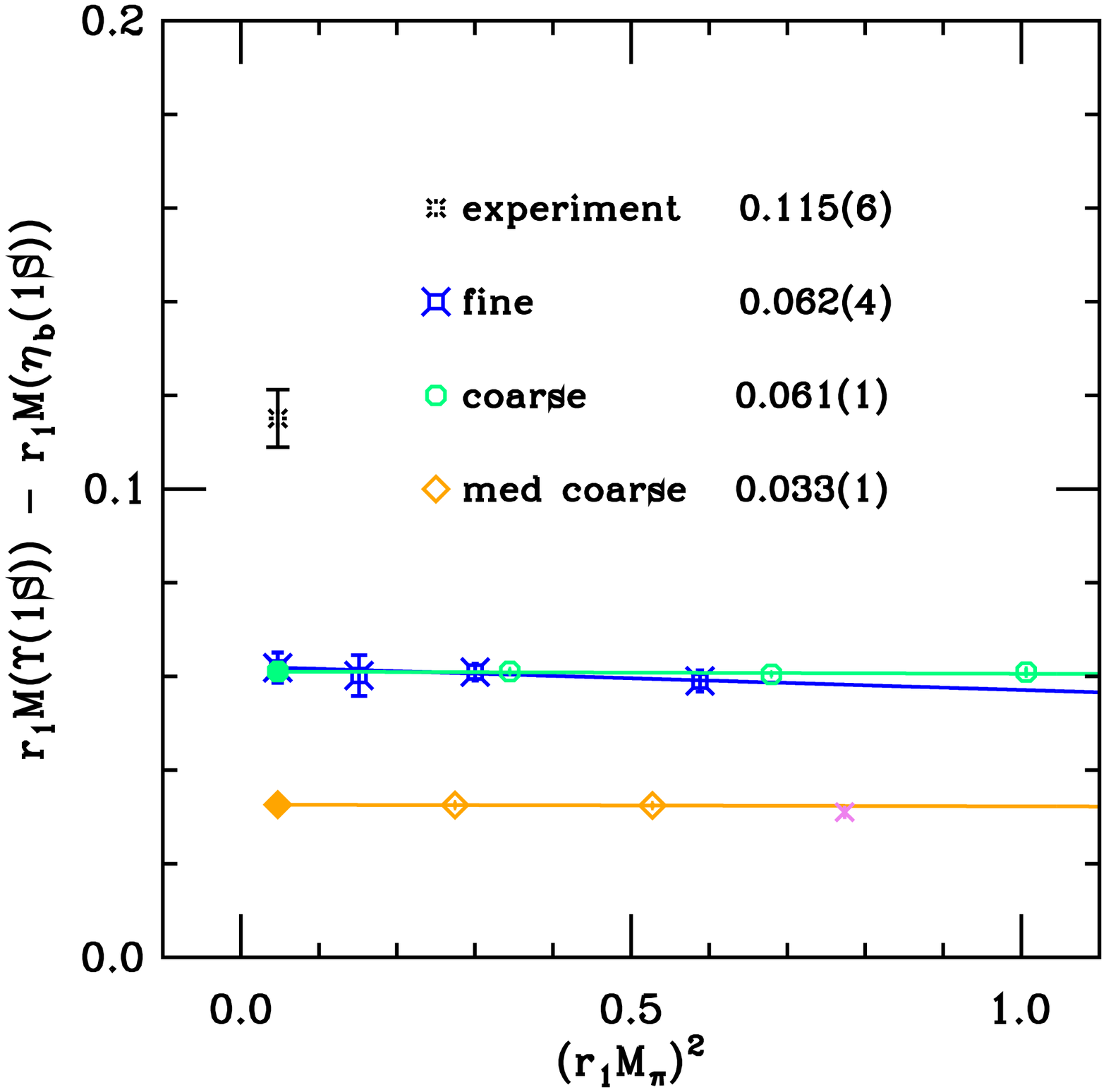}\\
  \end{minipage}
&
  \begin{minipage}[t]{0.5\textwidth}
\includegraphics[width=\figwidth]{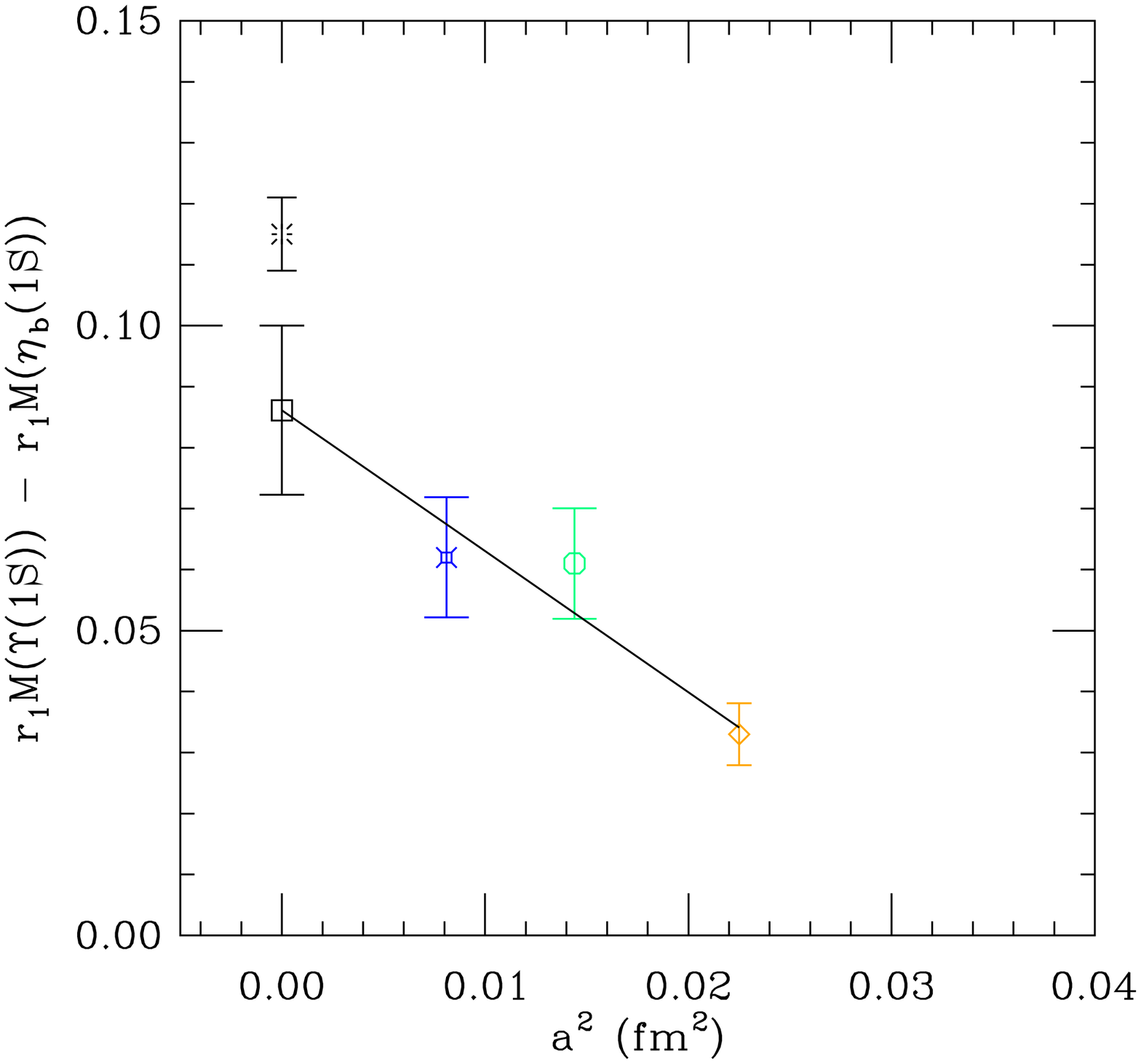}\\
  \end{minipage}
\end{tabular}
\end{center}
  \caption{The left panel shows the chiral extrapolation with only
    statistical errors shown.  The right panel shows the continuum
    extrapolation in $a^2$ with kappa tuning errors of 15\% included,
    resulting in 53(9) MeV.}
\label{fig:bhfs}
\end{figure}

\subsection{Bottomonium hyperfine splitting}  

In Fig.~\ref{fig:bhfs} we show results for hyperfine splitting of the
bottomonium ground state.  The continuum extrapolation gives 53(8)
MeV.  The $\eta_b$ was recently found \cite{Babar:2008,Babar:2009}
with a splitting of 71(4) MeV from the $\Upsilon(1S)$.  The HPQCD
collaboration reports 61(4)(13) \cite{Gray:2005} using an NRQCD method
with a chromomagnetic interaction of a quality comparable to ours.

\begin{figure}[t]
\begin{center}
\begin{tabular}{cc}
  \begin{minipage}[t]{0.5\textwidth}
\includegraphics[width=\figwidth]{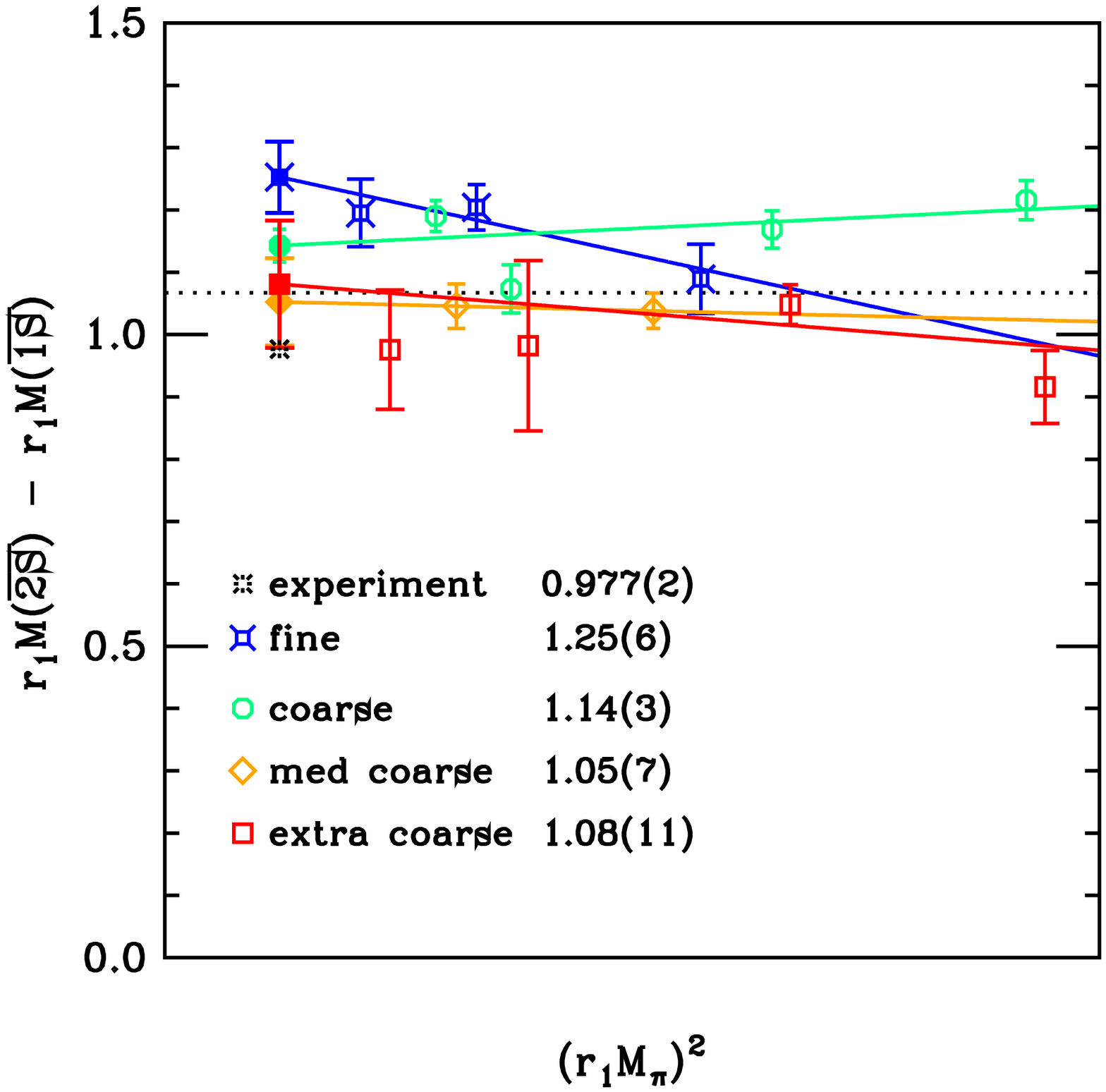}\\
  \end{minipage}
&
  \begin{minipage}[t]{0.5\textwidth}
\includegraphics[width=\figwidth]{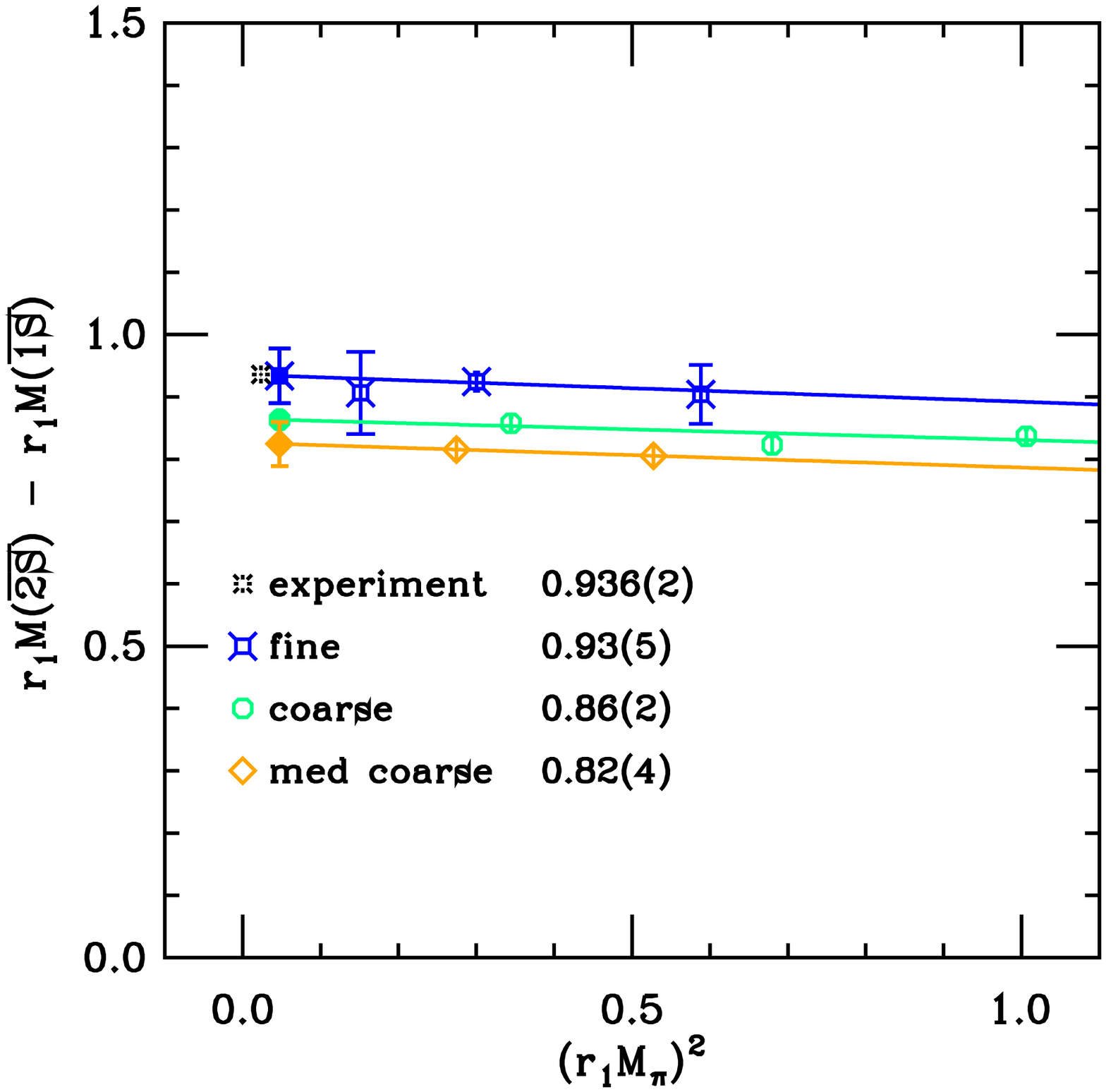}\\
  \end{minipage}
\end{tabular}
\end{center}
  \caption{Splitting of the spin-averaged $2S$ and $1S$ levels in
    charmonium (left) and bottomonium (right).  The dashed line
    indicates the physical open charm threshold. Since the
    $\eta_b^\prime$ has not been observed the ``experimental'' point
    uses only the $\Upsilon(2S)$ in the splitting.}
\label{fig:2S1Ssplitting}
\end{figure}

\subsection{$2S-1S$ level splitting}  

In Fig.~\ref{fig:2S1Ssplitting} we show results for the splitting of
the spin-averaged $\bm{\overline{2S}}$ and $\bm{\overline{1S}}$
levels. This quantity tests the ``central'' part of the quarkonium
effective potential.  We see that agreement with experiment in the
charmonium case is not good.  It is better in the bottomonium
case. Our fit model does not include open charm states.  So the $2S$
charmonium state could be confused with the nearby open charm
threshold that comes closer as the light sea quark mass decreases. The
dashed line locates the physical open charm threshold.  For the
bottomonium case the open bottom threshold is safely off scale.

\begin{figure}[t]
\begin{center}
\begin{tabular}{cc}
  \begin{minipage}[t]{0.5\textwidth}
\includegraphics[width=\figwidth]{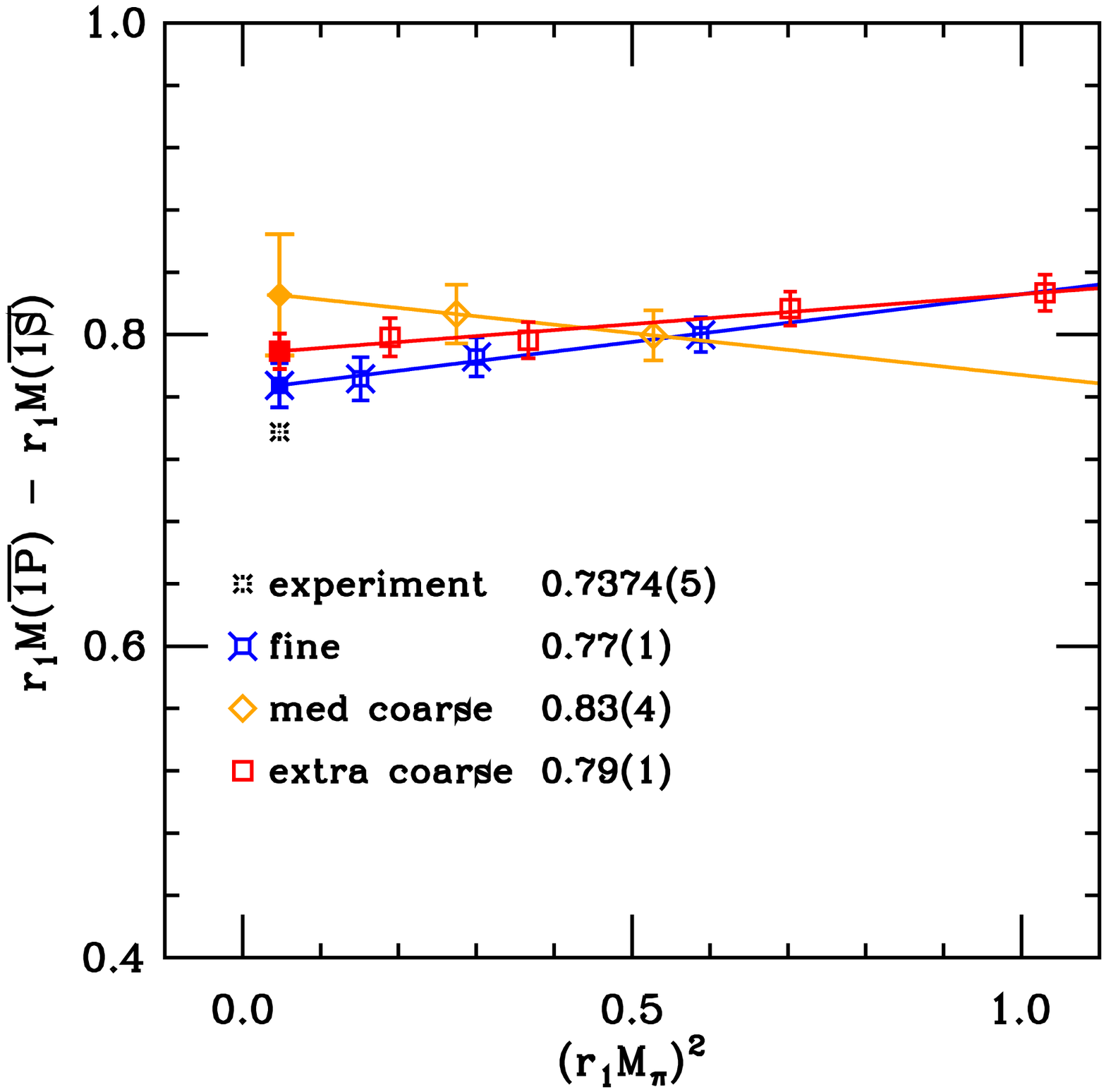}\\
  \end{minipage}
&
  \begin{minipage}[t]{0.5\textwidth}
\includegraphics[width=\figwidth]{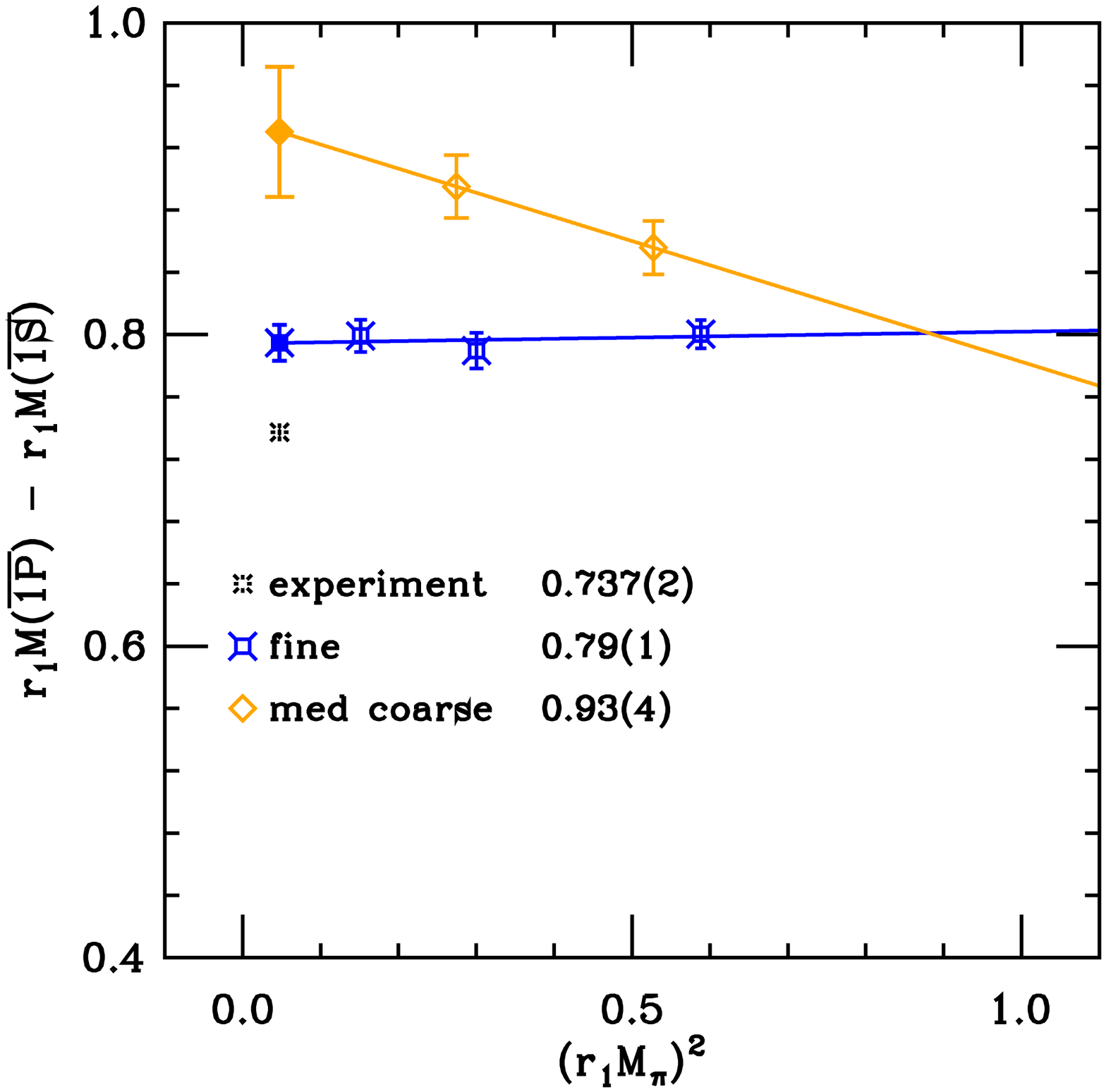}\\
  \end{minipage}
\end{tabular}
\end{center}
  \caption{Splitting of the spin-averaged $1P$ and $1S$ levels in
charmonium (left) and bottomonium (right).}
\label{fig:PSsplitting}
\end{figure}

\subsection{$1P-1S$ splitting}

The spin-averaged $\bm{\overline{1P}-\overline{1S}}$ splitting, shown
in Fig.~\ref{fig:PSsplitting}, also tests the central part of the
potential.  Within errors, our results seem to approach the
experimental value.

\begin{figure}[t]
\begin{center}
\begin{tabular}{cc}
  \begin{minipage}[t]{0.5\textwidth}
\includegraphics[width=\figwidth]{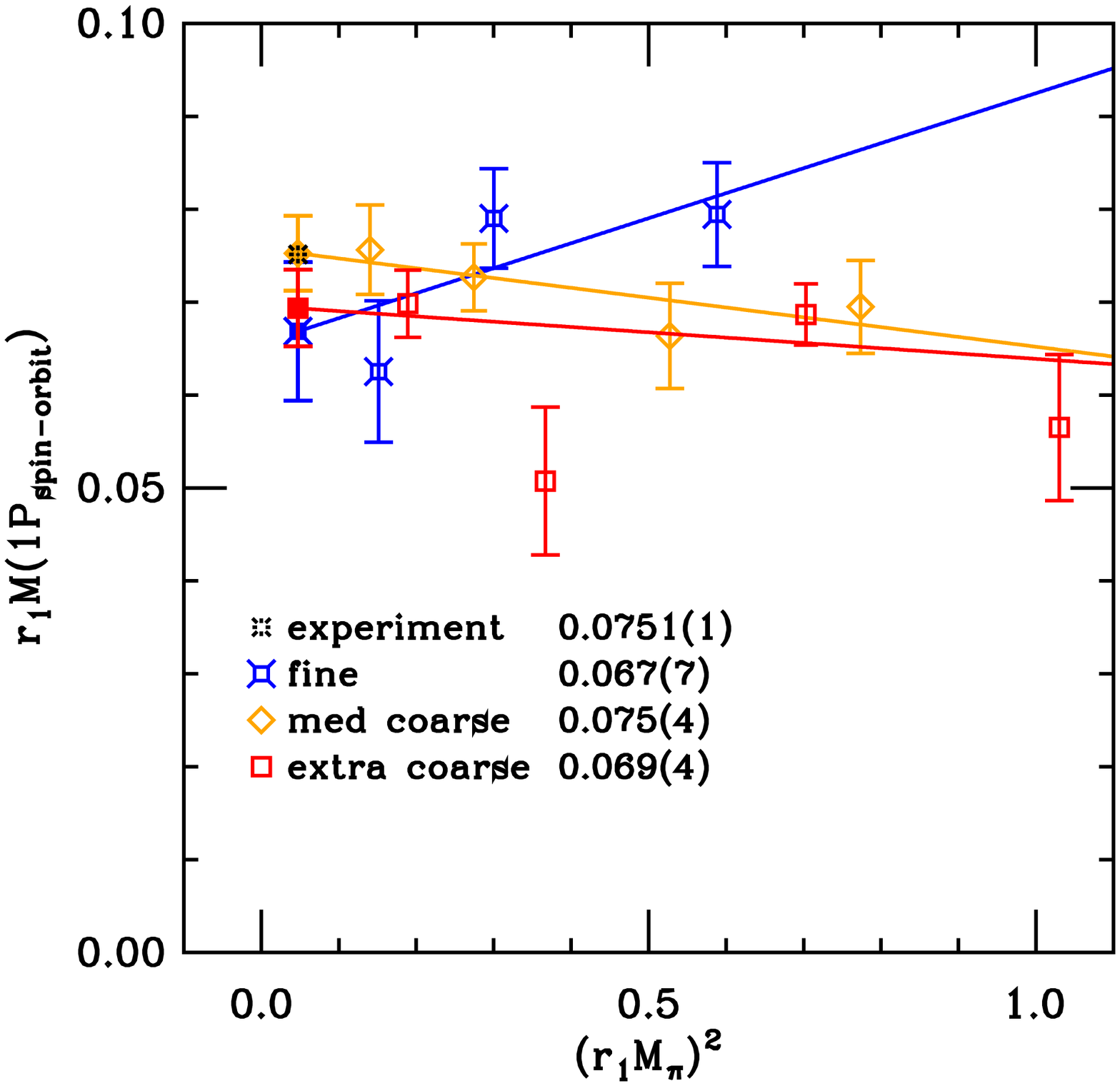}\\
  \end{minipage}
&
  \begin{minipage}[t]{0.5\textwidth}
\includegraphics[width=\figwidth]{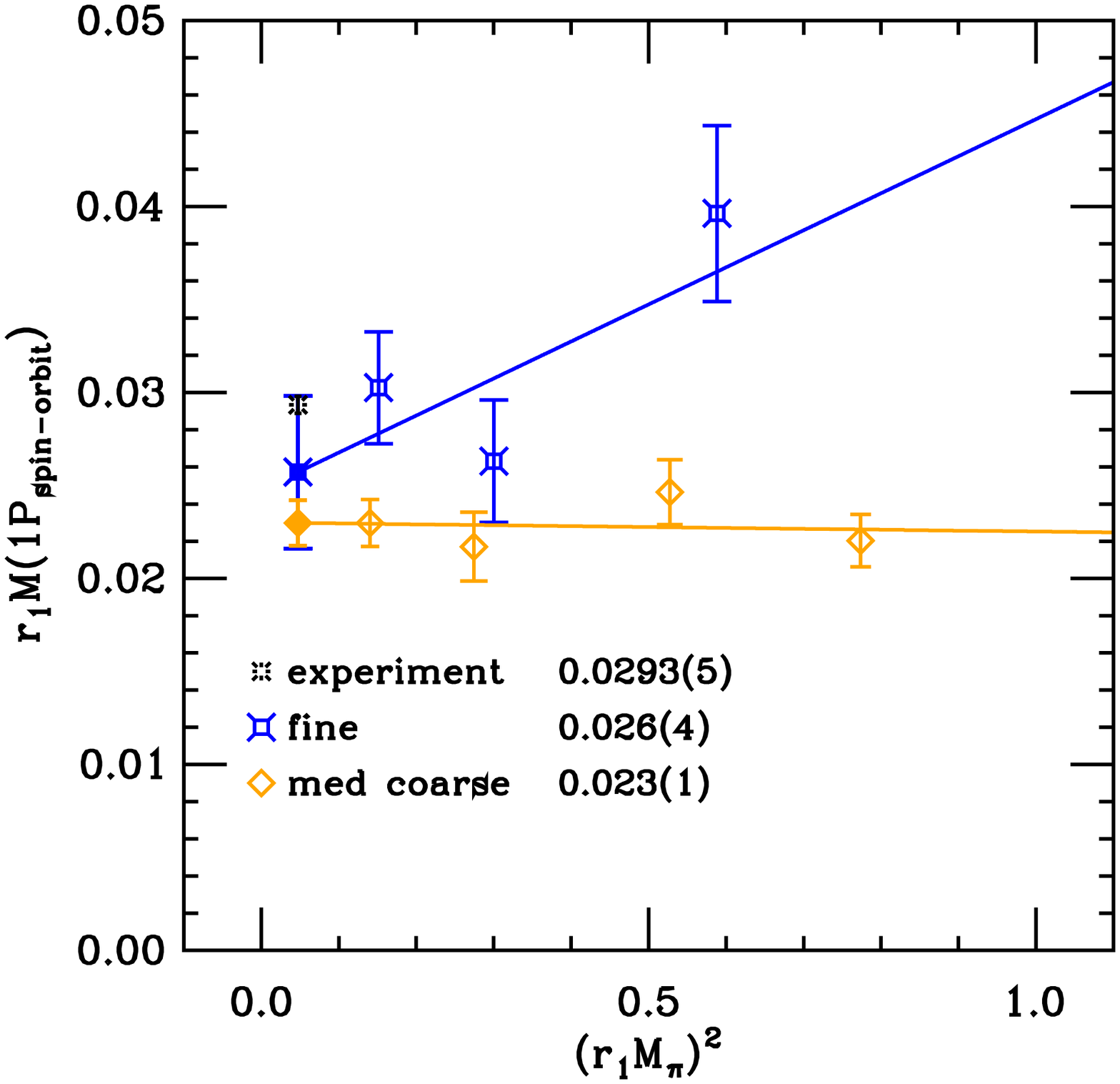}\\
  \end{minipage}
\end{tabular}
\end{center}
  \caption{Spin-orbit combination from the $1P$ levels for charmonium
    (left) and bottomonium (right).}
\label{fig:SO}
\end{figure}

\subsection{Spin-orbit and tensor components}

The contribution to the $J = 0$, 1, and 2 P-wave masses from the
spin-orbit term in the quarkonium effective potential can be isolated
with the combination
\begin{displaymath}
m_{1P_{\rm spin-orbit}} = 
  \case{1}{9}\left( 5 m_{c2} - 2 m_{c0} - 3 m_{c1}\right)
\end{displaymath}
Our result is shown in Fig.~\ref{fig:SO}.  This term tests the
strength of the chromoelectric interaction.  In both cases the results
seem to approach the experimental value in the chiral and continuum
limits.

\begin{figure}[t]
\begin{center}
\begin{tabular}{cc}
  \begin{minipage}[t]{0.5\textwidth}
\includegraphics[width=\figwidth]{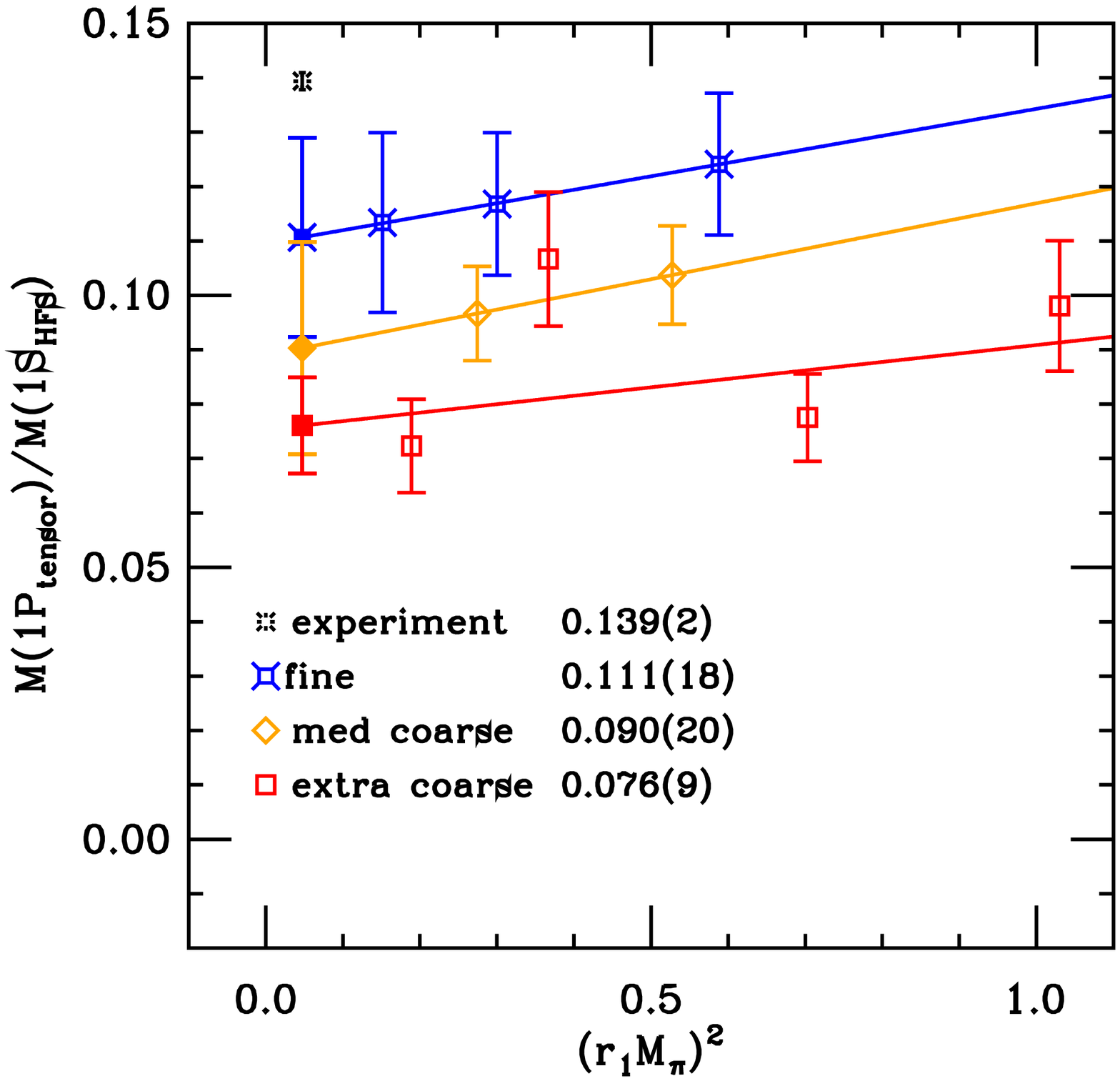}\\
  \end{minipage}
&
  \begin{minipage}[t]{0.5\textwidth}
\includegraphics[width=\figwidth]{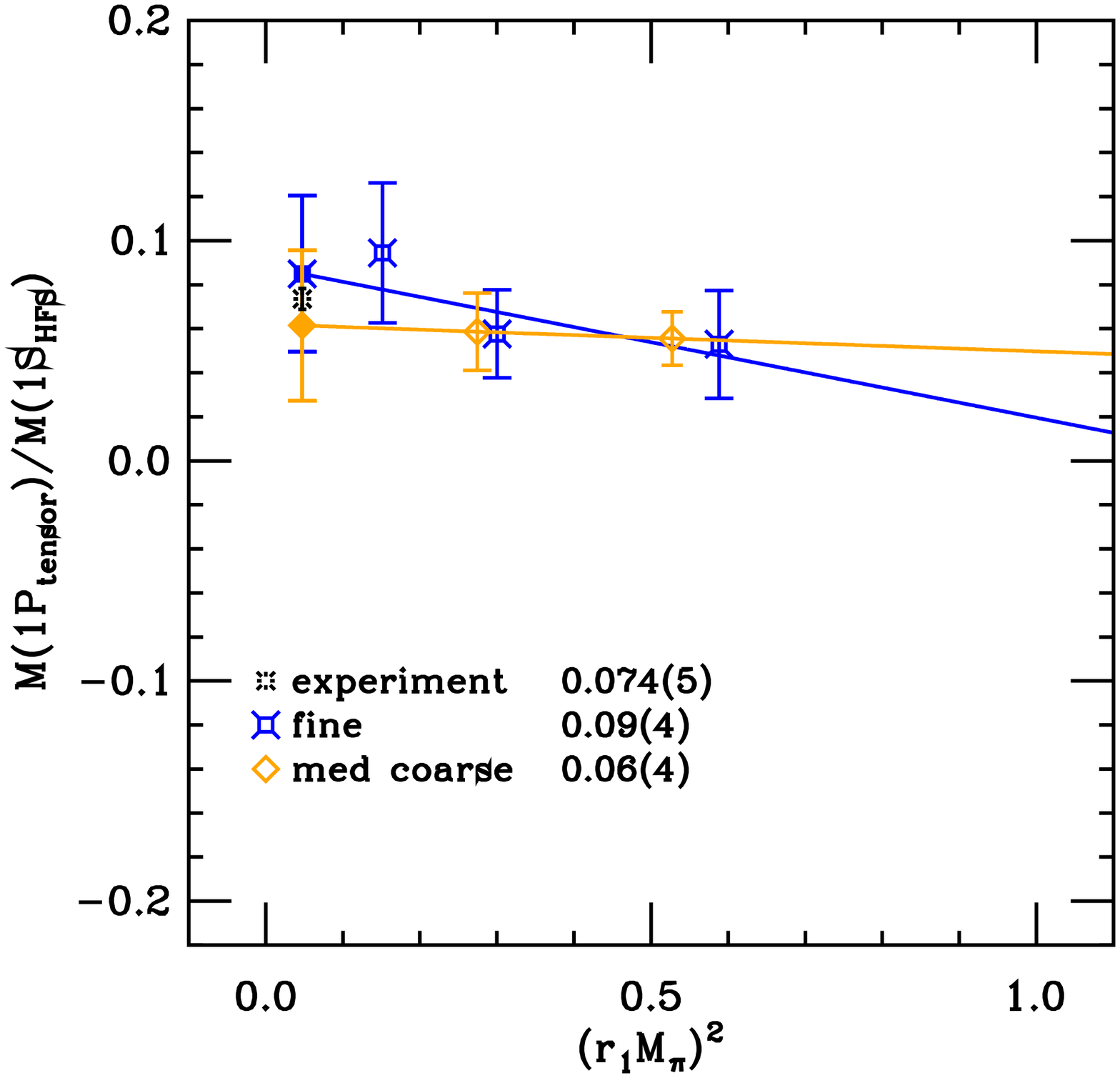}\\
  \end{minipage}
\end{tabular}
\end{center}
  \caption{The $1P$ tensor combination, divided by the $1S$ hyperfine
    splitting for charmonium (left) and bottomonium (right).}
\label{fig:tensor}
\end{figure}

Similarly, the contribution to the P-wave levels from the tensor component 
is proportional to the combination
\begin{displaymath}
m_{1P_{\rm tensor}} =
\case{1}{9}\left( 3 m_{c1} - m_{c2} - 2 m_{c0}\right),
\end{displaymath}
shown in Fig.~\ref{fig:tensor}.  Since the tensor and spin-spin
components both measure the strength of the chromomagnetic
interaction, here we divide by the 1S hyperfine splitting to see
whether they are proportional.  It appears that they are not.  Still
the results seem to approach the experimental values in the chiral and
continuum limits.
\subsection{Full spectrum}

\begin{figure}[t]
\begin{center}
\begin{tabular}{cc}
  \begin{minipage}[t]{0.5\textwidth}
\includegraphics[width=0.9\textwidth]{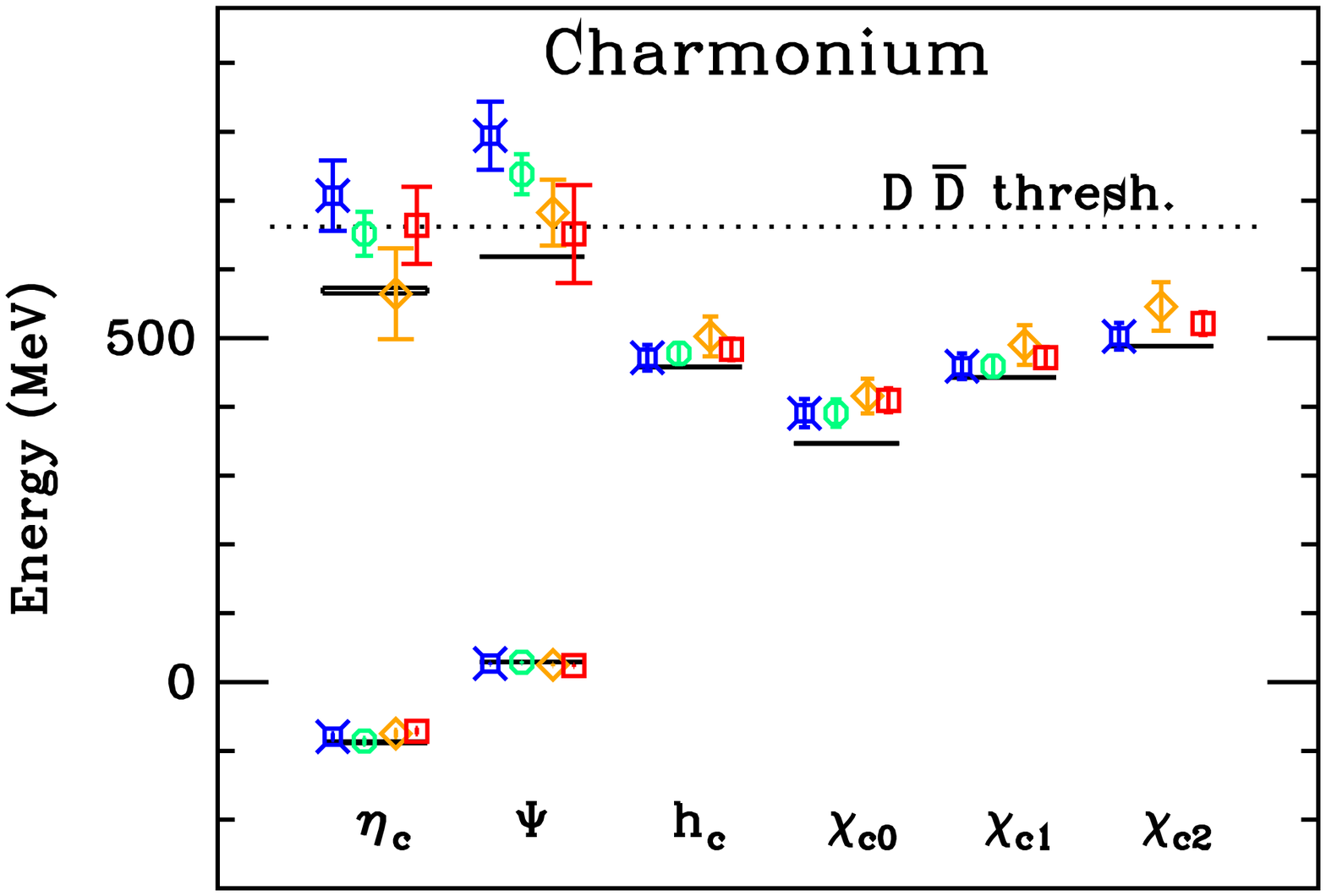}\\
  \end{minipage}
&
  \begin{minipage}[t]{0.5\textwidth}
\includegraphics[width=0.9\textwidth]{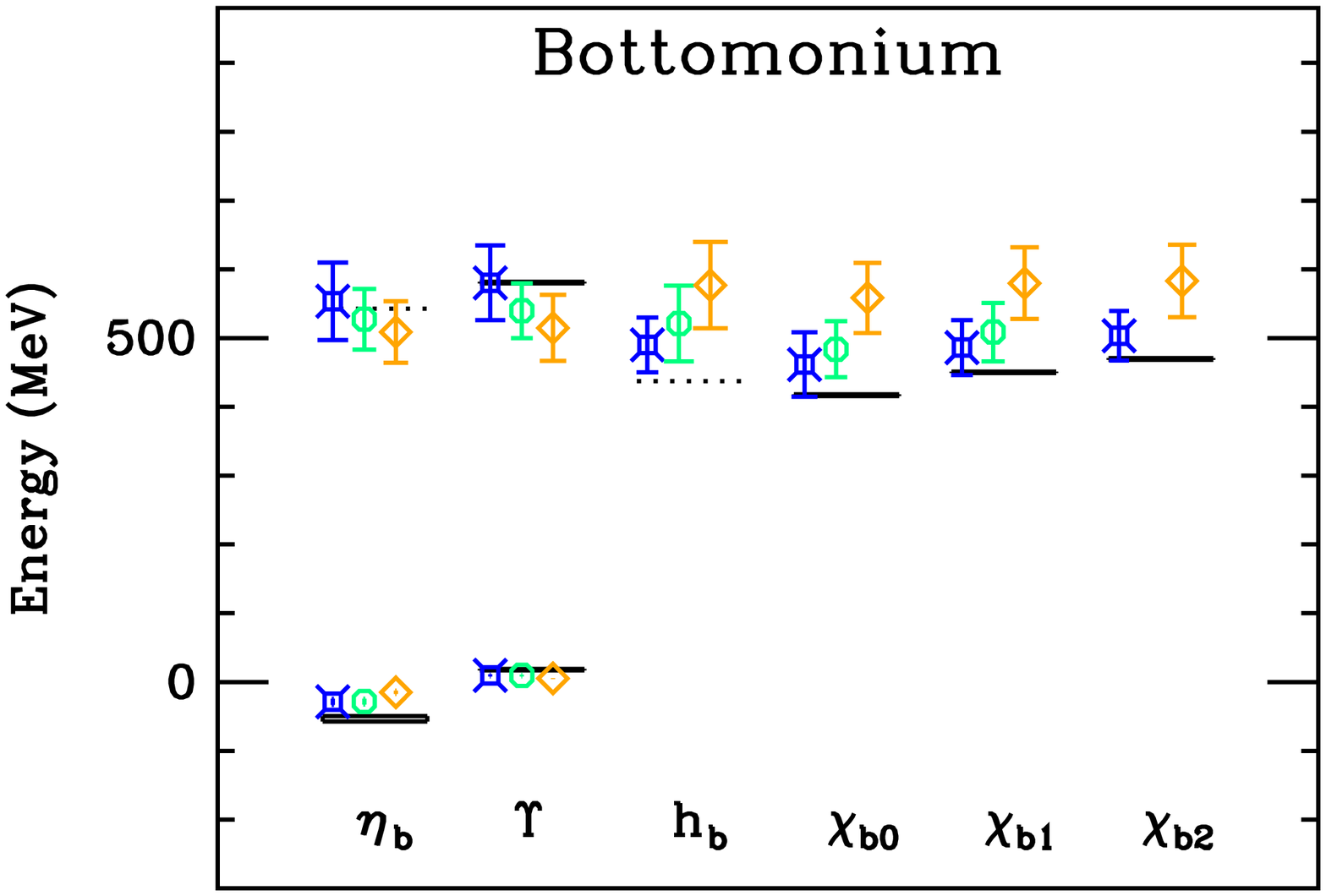}\\
  \end{minipage}
\end{tabular}
\end{center}
  \caption{All quarkonium levels in this study, constructed from
    splittings from the physical $\overline{1S}$~level for charmonium
    (left) and bottomonium (right).  Error bars include kappa tuning
    errors. Symbol colors distinguish the lattice spacings: 0.18
    fm (red), 0.15 fm (orange), 0.12 fm (green), 0.09 fm (blue)}
\label{fig:overview}
\end{figure}

In Fig.~\ref{fig:overview} we reconstruct the low-lying quarkonium
spectrum from splittings, starting from the experimental value for the
spin-averaged $1S$ level.

\section{Conclusion and Outlook}

We have seen that in most cases quarkonium level splittings are quite
insensitive to the light sea quark masses.  Systematic uncertainties
in tuning the quark masses are much larger than our statistical
errors.  With the present set of lattice spacings and the present
level of precision, the Fermilab action seems to perform well in the
charmonium system, but there are indications that lattice
discretization artifacts affect some of our bottomonium splittings.
Work currently underway seeks a more precise determination of the
charm and bottom masses and will use the still finer
MILC-collaboration $0.06$ fm lattices.

\acknowledgments

Work is supported by grants from the US Department of Energy and US
National Science Foundation.  The lattice ensembles used in this study
were generated by the MILC collaboration.  Computations for this work
were carried out on facilities of the USQCD Collaboration, which are
funded by the Office of Science of the U.S. Department of Energy. T.B.
acknowledges current support by the DFG (SFB/TR55).


\providecommand{\href}[2]{#2}\begingroup\raggedright\endgroup

\end{document}